\documentclass[final,2p,times,twocolumn]{elsarticle}
\usepackage{epsfig}

\usepackage{amsfonts}
\usepackage{mathrsfs}
\usepackage{bm}

\biboptions{sort&compress}

\usepackage[usenames,dvipsnames]{xcolor}

\definecolor{R}{rgb}{0.9,0.0,0.1}
\definecolor{G}{rgb}{0.3,0.7,0.345}
\definecolor{Y}{RGB}{255,255,0}

\makeatletter
\def\ps@pprintTitle{%
  \let\@oddhead\@empty
  \let\@evenhead\@empty
  \def\@oddfoot{\reset@font\hfil\thepage\hfil}
  \let\@evenfoot\@oddfoot
}
\makeatother

\def\be{\begin{equation}}
\def\ee{\end{equation}}
\def\bea{\begin{eqnarray}}
\def\eea{\end{eqnarray}}
\def\bfl{\begin{flushleft}}
\def\efl{\end{flushleft}}
\def\bfr{\begin{flushright}}
\def\efr{\end{flushright}}
\def\bc{\begin{center}}
\def\ec{\end{center}}
\def\ben{\begin{enumerate}}
\def\een{\end{enumerate}}
\def\bit{\begin{itemize}}
\def\eit{\end{itemize}}

\def\dzn{,\kern-0.1em,}

\journal{Communications in Nonlinear Science and Numerical Simulation}

\begin{document}

\begin{frontmatter}

 \title{Application of largest Lyapunov exponent analysis on the studies of dynamics under external forces\tnoteref{label1}}

\title{ Application of largest Lyapunov exponent analysis on the studies of dynamics under external forces}

\author[rvt]{Jovan Odavi\' c}
\author[pmf]{Petar Mali \corref{cor1}}
\ead{petar.mali@df.uns.ac.rs}
\author[focal]{Jasmina Teki\' c}
\author[pmf]{Milan Panti\' c}
\author[pmf]{Milica Pavkov - Hrvojevi\' c}

\address[rvt]{Institut f\" ur Theorie der Statistishen Physik - RWTH Aachen University,
Peter-Gr\"unberg Institut and Institute for Advanced Simulation, Forschungszentrum J\"ulich, Germany}
\address[pmf]{Department of Physics, Faculty of Sciences, University of Novi Sad,
Trg Dositeja Obradovi\' ca 4, Novi Sad, Serbia}
\address[focal]{"Vin\v ca" Institute of Nuclear Sciences,
Laboratory for Theoretical and Condensed Matter Physics - 020,
University of Belgrade, PO Box 522, 11001 Belgrade, Serbia}

\cortext[cor]{Corresponding author}

\begin{abstract}
Dynamics of driven dissipative Frenkel-Kontorova model is examined
by using largest Lyapunov exponent computational technique.
Obtained results show that besides the usual way where behavior of the system
in the presence of external forces is studied by analyzing its
dynamical response function, the largest Lyapunov exponent analysis
can represent a very convenient tool to examine system dynamics.
In the dc driven systems, the critical depinning force for particular structure could
be estimated by computing the largest Lyapunov exponent.
In the dc+ac driven systems, if the substrate potential is the standard
sinusoidal one, calculation of the largest Lyapunov exponent offers
a more sensitive way to detect the presence of Shapiro steps.
When the amplitude of the ac force is varied the behavior of the largest
Lyapunov exponent in the pinned regime completely reflects the
behavior of Shapiro steps and the critical depinning force, in
particular, it represents the mirror image of the amplitude
dependence of critical depinning force.
This points out an advantage of this technique since by calculating the
largest Lyapunov exponent in the pinned regime we can get an insight
into the dynamics of the system when driving forces are applied.

\end{abstract}
\end{frontmatter}

\section{Introduction}
\label{intro}

The Frenkel - Kontorova (FK) model is widely used to describe systems where competition between length scales determines the ground state energy.
Dissipative FK model has been often used as one of the most suitable models for description of different kinds of phenomena in many fields of physics,
such as charge or spin density wave systems \cite{Grun,Zettl,Thorn, ThLyoI, ThLyoII, Kriza}, vortex matter \cite{Kokub, Kol},
Josephson-junction arrays biased by external currents \cite{Benz, Sell, Free,Medvedeva,Shukrinov} and in recent years
even superconducting nanowires \cite{Dins, Bae}.
The one-dimensional FK model is a simple classical model which describes
a chain of particles, usually identical, coupled to their nearest neighbors and subjected to
the periodic or quasiperiodic \cite{kvazi,kvazi2} on-site substrate potential.
In particular, in the standard FK model, interaction between particles is harmonic, and the on-site substrate potential is sinusoidal.
Although very simple, the one-dimensional standard FK model
exhibits rich dynamics when it is subjected to external driving forces \cite{OBBook, Monograph}.

The FK model has two distinct types of system ground states, commensurate and incommensurate ones, 
(in the commensurate grounds states interparticle average distance is a rational number, while for the incommensurate ones
it is irrational). 
If an external dc driving force is applied (the dc driven FK model), there exists a critical threshold value
i.e. the critical depinning force $F_{\mathsf{c}}$ which separates two different dynamical regimes: the pinned and the sliding regimes.
The type of the structure, whether it is commensurate or incommensurate determines depinning transition.
In the commensurate structures, particles exhibit the motion with stick and slip intervals or
intermittent behavior (Intermittencies type I) \cite{OBBook, Monograph}.
On the other hand, in incommensurate structures, the depinning appears only if the system is above  Aubry transition while
if it is below, there is no depinning transition, i.e. under any force $F>0$ particles are in sliding regime \cite{OBBook, Monograph,Flor}.

In the case of dc+ac driven FK model, the dynamics
is characterized by the appearance of the staircase macroscopic response
or Shapiro steps in the response function $\bar{\upsilon}(\bar F)$ of the system \cite{Flor,  Falo, FlorAP}.
These steps appear due to interference or dynamical mode-locking of the internal frequency (that comes from
the motion of particles over periodic substrate potential) with the frequency of external ac force.

In the case of incommensurate structures, the ac-driven dissipative dynamics exhibits the dynamical Aubry
transition \cite{OBBook,Flor,FlorAP,Monograph}.
The dynamical hull function that describes the driven structure becomes nonanalytical
above the transition point and the result of this is the dynamical
locking $\bar{\upsilon}(\bar F)$ at certain resonant values.
The dynamical mode locking for the commensurate
and incommensurate structures, respectively, appear to be one of the universal
features of the systems with the competition of time scales in the ac-driven dynamics.

Critical depinning force and the Shapiro steps have been the subjects of our previous investigations \cite{Tekic,Monograph,Tekic2,Tekic3,Mali}.
The standard way to examine dynamics of a driven system is by analyzing the response function.
However, this analysis could be very difficult since in the response function of the systems, Shapiro steps particularly the subharmonic ones
are often hardly visible.
In realistic systems such as the CDW systems and the systems of Josephson-junction arrays, particularly in the experimental examinations,
instead of analyzing the current-voltage characteristics where Shapiro steps could be hardly detected, examination of differential resistance
provides more accurate way to determine the presence and properties of the Shapiro steps \cite{Grun,Zettl,Thorn,Benz}.
In general, in studies of ac-driven systems, the main interest is always focused on the existence and robustness (structural stability) of
the resonant solutions against the varying of the system parameters, and
therefore, an accurate detection of Shapiro steps have always been  of great importance for both theoretical and experimental studies.

In this paper, we will show that dynamics of the dc+ac driven system could be examined in detail and just as successfully
by using the largest Lyapunov exponent computation technique.
The largest Lyapunov exponent not only reveals the existence of Shapiro steps but
also provides an insight into dynamical response of the system and properties of the Shapiro steps.
The paper is organized as follows: The model is introduced
in Sec. II, simulation results are presented in Sec. III.
Finally, Sec. IV concludes the paper.

\section{Model}
\label{model}

\begin{figure}[ht] 
\includegraphics[width=5.0cm]{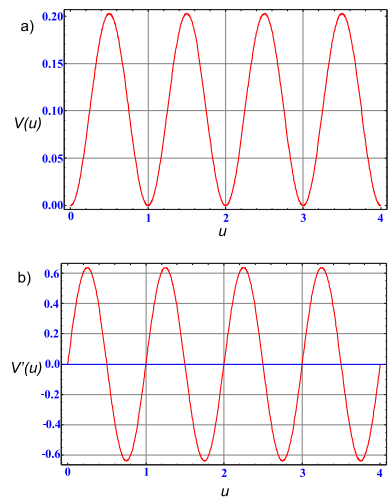}
\centering
\caption{\label{Fig1}
a) Sinusoidal potential described in Eq(\ref{sp}) for $K=4$; b) Its corresponding force $F(u)=-\frac{\mathsf{d}V}{\mathsf{d}u}$.}
\end{figure}

We consider the dissipative (overdamped) dynamics of a series of coupled harmonic
oscillators $u_l$ subjected to the standard sinusoidal substrate potential:
\begin{equation}
\label{sp}
V(u)=\frac {K}{(2\pi )^2}[1-\cos (2\pi u)],
\end{equation}
The sinusoidal potential is usually written in this form so that
$K$ is the pinning strength and the period is $1$, $V(u+1)=V(u)$ \cite{OBBook, FlorAP,Monograph}.
The potential $V(u)$ given in Eq. (\ref{sp}) and the corresponding force $F(u)=-\frac{\mathsf{d}V}{\mathsf{d}u}$ are presented in Figure \ref{Fig1}.

The system is driven by dc and ac forces:
\begin{equation}
\label{F}
F(t) = F_{\mathsf{dc}} +F_{\mathsf{ac}}\cos (2\pi \nu _{0}t).
\end{equation}
where $F_{\mathsf{dc}}$ is the dc force, while $F_{\mathsf{ac}}$ and $2\pi\nu_0$ are the amplitude and the circular frequency of the ac force, respectively.
Since the ac force $F_{\mathsf{ac}}\cos (2\pi\nu_0 t)$ drives the system of particles forward and back during the time of one period $\tau=\frac{1}{\nu_0}$,
the average driving force is $\bar{F}=F_{\mathsf{dc}}$.

The equations of motion are:
\begin{equation}
\label{u}
\dot u_l
=u_{l+1}+u_{l-1}-2u_{l}+\frac{K}{2\pi}\sin (2\pi u_l)+F(t),
\label{eq} \end{equation}
where $l=-\frac N2,...,\frac N2$, $u_l$ is the position of $l$th particle.
This system of equations is normalized in such a way that the parameters which characterize overdamped dynamics are the strength of potential ($K$) and
characteristics of the applied force ($F_{\mathsf{ac}}$ and $\nu_0$) \cite{OBBook, FlorAP,Monograph}.

When the system is driven by the forces given in Eq. (\ref{F}), two different frequency scales appear in the system:
the frequency of the external periodic force $\nu_0$ and the characteristic frequency of the motion of particles
over the sinusoidal potential driven by the average force $F_{\mathsf{dc}}$.
The competition between those frequency scales can result in the appearance of resonance (dynamical mode-locking or Shapiro steps).
The steps are called harmonic if the locking appears for integer values of frequency,
whereas if it appears for the rational noninteger values of the frequency they are called subharmonic.
If $\{u_l(t)\}$ is steady state solution of (\ref{u}), the transformation
\begin{equation}
\label{sim}
\sigma_{i,j,m}\{u_l(t)\}=\{u_{l+i}(t-\frac{m}{\nu_0})+j\}=\{u_l'(t)\}
\end{equation}
produces another steady state solution, where $i,j,m$ are integers.

If  solution is invariant under a symmetry operation (\ref{sim})
\begin{equation}
\sigma_{i,j,m}\{u_l(t)\}=\{u_l(t)\}
\end{equation}
then solution of the system (\ref{u}) is called resonant and its (particle and time) average velocity $\bar{\upsilon}$ satisfies the following relation:
\begin{equation}
\label{vel}
\bar{\upsilon} = \frac {i\omega + j}{m}\nu _0,
\end{equation}
$m=1$ for harmonic, and $m>1$ for subharmonic  steps, while
$\omega=\langle u_{l+1}-u_l \rangle$ is the average interparticle distance also known as the winding number
($\omega$ is rational for the commensurate structures and irrational for the incommensurate ones).
The triplet $i,j$, and $m$ is unique only for incommensurate structures \cite{OBBook}.

The equations of motion (\ref{eq}) have been integrated using
the fourth order Runge-Kutta method with periodic boundary conditions.
The time step used in the simulation was $0.01 \tau$, where $\tau=\frac{1}{\nu_0}$.
The force is varied adiabatically with the step $\Delta F_{\mathsf{dc}} = 1 \times 10^{-5}$.

We will focus on the calculation of the largest Lyapunov exponent, which we denote by $\lambda$.
Detailed explanation of our approach and numerical procedure can be found in Ref. \cite{Odavic}.
In further text, we will refer to the quantity of largest Lyapunov exponent as the Lyapunov exponent (LE) for convenience.
The LE will be analyzed in different regimes and for different parameters of the systems.
It is well known that the LE can indicate the chaotic behavior in dynamical systems, and therefore,
the presented studies will give us also an information about the presence or absence of chaos in the driven dissipative one-dimensional FK model.

\section{Results}
\label{results}
We present a comparative study of system dynamics under external forces using the response function and LE as a function of average driving force.

\subsection{dc driven system}
\label{dc}

\begin{figure*}[ht] 
\includegraphics[width=15.0cm]{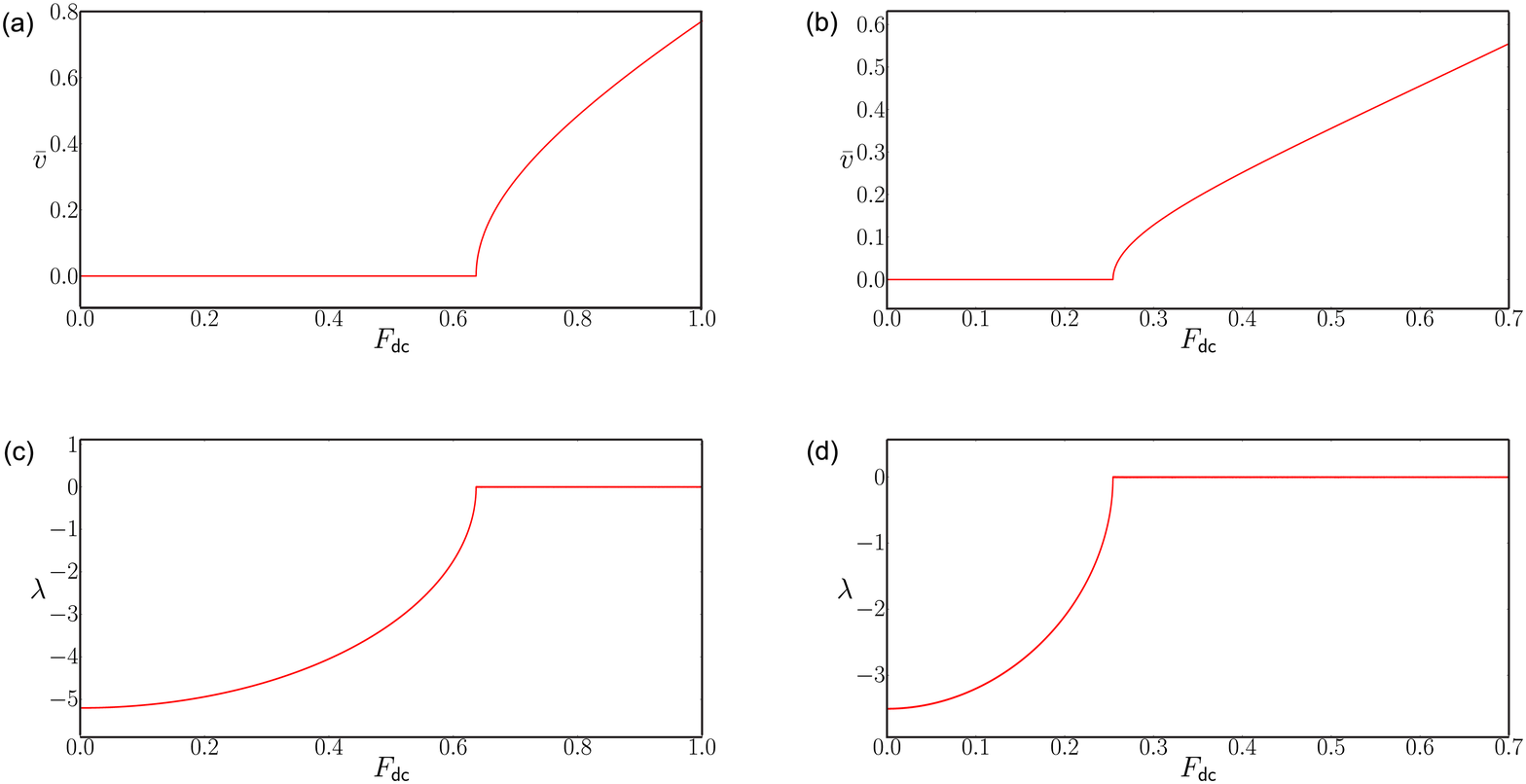}
\centering
\caption{\label{Fig2}
Average velocity as a function of average driving force $\bar{\upsilon}(F_{\mathsf{dc}})$ and
Lyapunov exponent as a function of average driving force $\lambda(F_{\mathsf{dc}})$ for $K=4$,
$F_{\mathsf{ac}}=0$ (dc driven system) and two different commensurate structures $\omega=1$ in (a) and (c) and $\omega=1/2$ in (b) and (d). }
\end{figure*}

First, we will consider the simplest case, the dc driven standard FK model.
In Figure \ref{Fig2}, the response function $\bar{\upsilon}(F_{\mathsf{dc}})$ and the Lyapunov exponent $\lambda(F_{\mathsf{dc}})$ are presented
for the same range of the applied dc force and two different commensurate structures.
A clear dynamical transition point between the pinned and the sliding regime can be observed.
Naturally, in the case of the response function $\bar{\upsilon}(F_{\mathsf{dc}})$, this transition happens when the average velocity reaches
the non-zero value $\bar{\upsilon} \neq 0$ for the lowest $F_{\mathsf{dc}}$ in Figure \ref{Fig2} (a) and (b),
while in the other case in Figure \ref{Fig2} (c) and (d), the transition is identified by the point where
LE becomes zero $\lambda =0$.
Given that the system is dc driven, the LE will keep its zero value in the whole sliding regime.
The critical depinning force in the case of $\omega = 1$ (Figure \ref{Fig2} (a)) approximately has the value as $\max ({V'(u)})$ (see Figure \ref{Fig1} (b)).
The system is more stable in the case of $\omega =1$ than in the case of $\omega=1/2$, which means that the force required to depin the system is larger.
This comes as no surprise given that in the first case, we have one particle, while in the second one,
we have two particles per potential well, and competing interactions between the particles makes the system less stable and therefore, easier to depin.
Influence of particular configuration (winding number) on the critical depinning force has been examined in our previous work \cite{MaliCNSNS},
where it can be seen that in the standard case, the system is the most robust for $\omega = 1$, while
as the number of particles per potential well increases (denominator of winding number increases) it becomes less stable.
An equivalent observation can be made from the Lyapunov exponent in this context.
We can say something about the stability of a particular configuration just by looking at the Lyapunov exponent for $F_{\mathsf{dc}} = 0$
when no dc driving force is applied.
For the same value of $F_{\mathsf{dc}}$, the LE is more negative in the case $\omega =1$ where $\lambda \approx -5$ than
in the case $\omega =1/2$ where $\lambda \approx -3.5$ .
As the LE gets more negative the configuration is going to be more robust to the external disturbances or
equivalently the more negative the LE is, the more quickly the trajectory returns to its unperturbed path.

\begin{figure}[ht] 
\includegraphics[width=8.0cm]{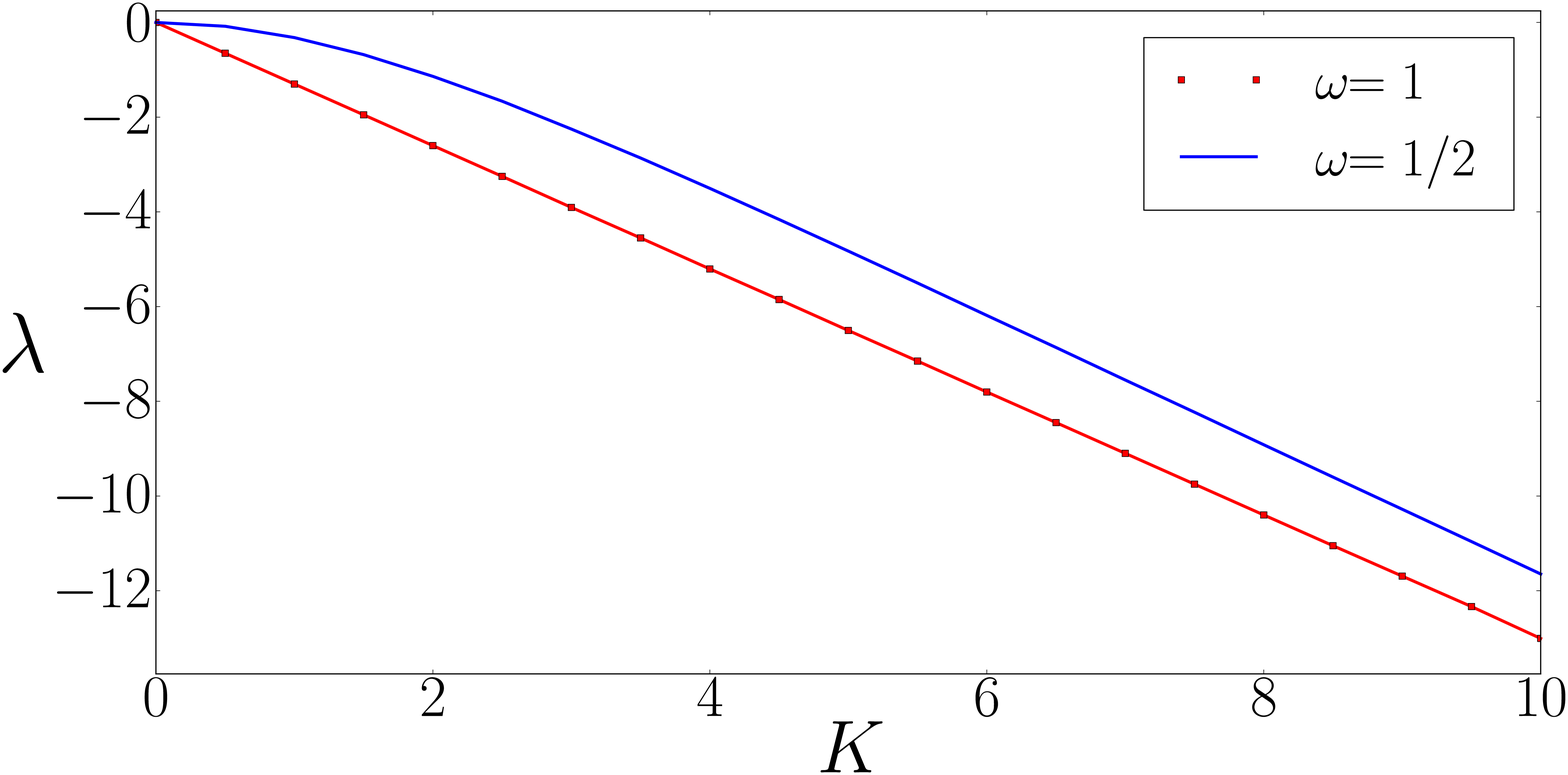}
\centering
\caption{\label{Fig3}
Lyapunov exponent as a function of pinning strength $K$ without driving force for (a) $\omega=1$ and (b) $\omega=1/2$. }
\end{figure}

If we consider only the static case (there is no driving force $\bar F=0$),
we can examine how different types of structures or potential strength influence the LE.
In Figure \ref{Fig3}, the Lyapunov exponent as a function of pinning strength for two different commensurate structures is presented.
As the pinning strength $K$ increases or if the commensurate structure is more stable such as the one for $\omega =1$, $\lambda $ becomes more negative.
This means that much higher force will have to be applied in order to set the particles into collective motion.

\subsection{dc+ac driven system}
\label{ac}

In the case of the dc+ac driven FK model, the dynamics is characterized by the appearance of Shapiro steps.
We will examine both the commensurate and incommensurate structures.

\begin{figure*}[ht] 
\includegraphics[width=13.0cm]{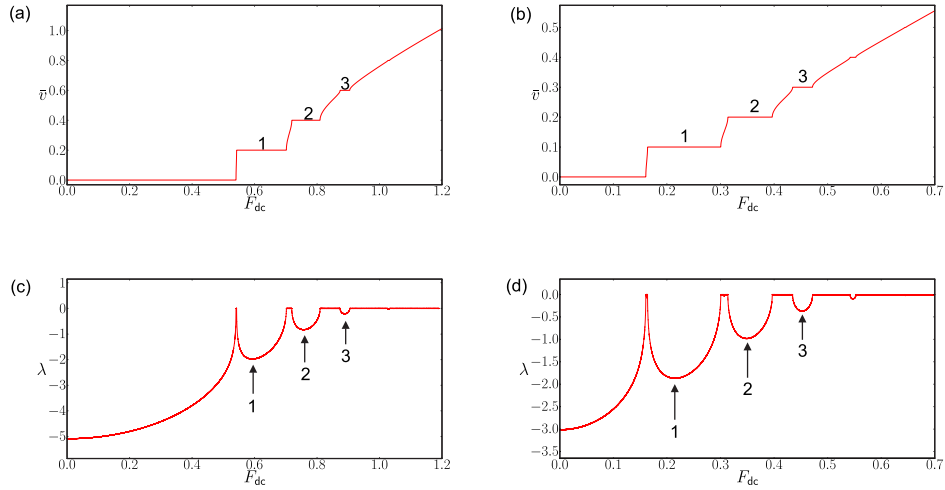}
\centering
\caption{\label{Fig4}
Average velocity (a) and (b) and  the Lyapunov exponent (c) and (d) as a function of average driving force for $K=4$,
$F_{\mathsf{ac}}=0.2$, $\nu_0=0.2$ and two different commensurate structures $\omega=1$ in (a) and (c) and $\omega=1/2$ in (b) and (d).
Numbers mark first, second and third harmonics. Triplets of integers $(i,j,m)$ in Eq. \ref{vel}
that determine the first three harmonic steps in case of $\omega=1$ are $(1,0,1)$, $(2,0,1)$, $(3,0,1)$ respectively and
in case $\omega=1/2$ $(1,0,1)$, $(2,0,1)$, $(3,0,1)$. These triplets are not unique in the case of commensurate structures.
}
\end{figure*}

In Figure \ref{Fig4}, the response function and the corresponding Lyapunov exponent as a function of average driving force
for two different commensurate structures are presented.
As we can see, in the pinned regime for the same $F_{\mathsf{dc}}$, LE is more negative  for more
stable structures such as $\omega=1$.
In the sliding regime, negative values of Lyapunov exponent correspond to
the Shapiro steps (the region of $F_{\mathsf{dc}}$ for which $\lambda
<0$ in Figure \ref{Fig4} (c) and (d) corresponds to the same region of
$F_{\mathsf{dc}}$ for which steps appear in Figure \ref{Fig4} (a) and (b)).

\begin{figure*}[ht] 
\includegraphics[width=17.0cm]{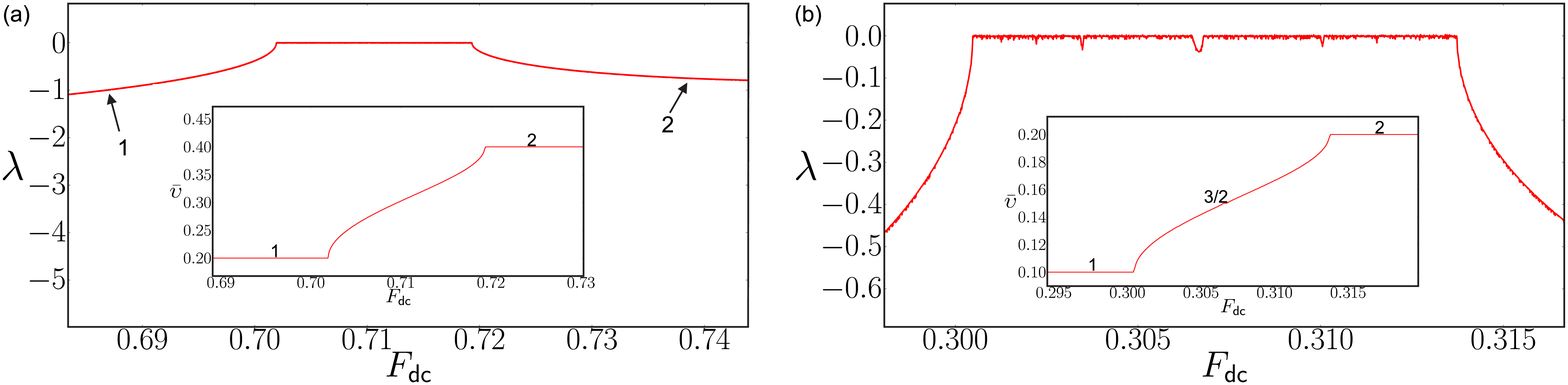}
\centering
\caption{\label{Fig5}
Lyapunov exponent as a function of average driving force between first and second harmonics
for $K=4$, $F_{\mathsf{ac}}=0.2$, $\nu_0=0.2$ and two different commensurate structures $\omega=1$ at (a) and $\omega=1/2$ at (b).
The insets show the corresponding average velocities.
}
\end{figure*}

Since this is the standard FK model, the subharmonic Shapiro steps appear only for noninteger, rational values of $\omega $,
while for the integer ones, only harmonic steps exist \cite{Renne}.
As we can see in Figure \ref{Fig4} (a) and (b), for both commensurate structures $\omega=1$ and $\omega=1/2$, only harmonic steps are visible
which might lead to conclusion that there are no subharmonic steps.
However, if we enlarge the region between the first and the second harmonics, results obtained for LE reveal the presence of subharmonic steps  for $\omega=1/2$.
In  Figure \ref{Fig5}, the enlarged regions between the first and second harmonic of the LE in Figure \ref{Fig4} (c) and (d) are presented.
For comparison, the corresponding response functions are given in the insets.

In the case of $\omega =1$, there is no subharmonic mode locking as Figure \ref{Fig5} (a) clearly shows.
When $\omega $ is noninteger in Figure \ref{Fig5} (b), only the halfinteger step $\frac 32$ appears in the response function,
while on the other hand, the Lyapunov exponent reveals a whole series of subharmonic steps.
It was shown in our previous work that those subharmonic steps appear according to Farey rule  \cite{Odavic}.
\begin{figure*}[ht] 
\includegraphics[width=12.0cm]{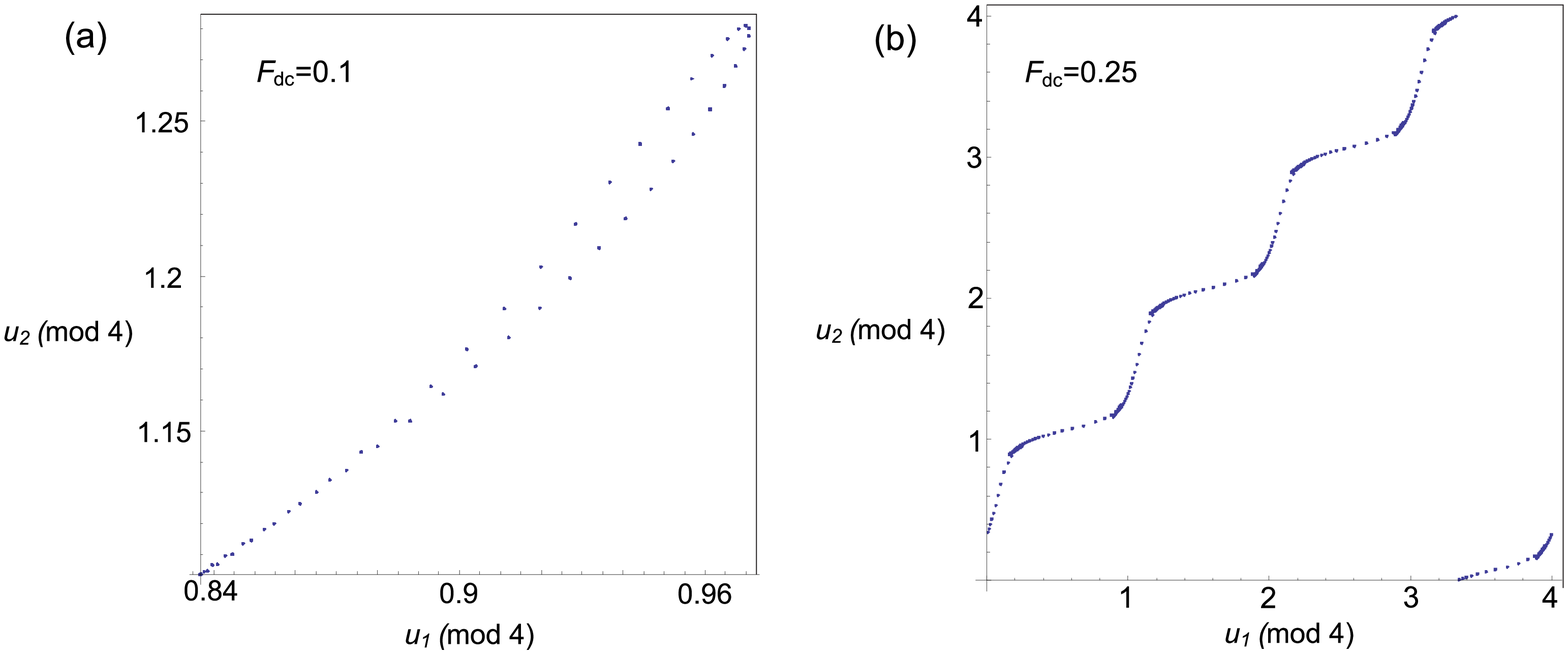}
\centering
\caption{\label{Fig6}
Particle motion for  $F_{\mathsf{ac}}=0.2$, $\nu_0=0.2$, in pinning $F_{\mathsf{dc}}=0.1$ (a), and sliding $F_{\mathsf{dc}}=0.25$ (b) regimes.}
\end{figure*}
In order to better understand the dynamics, let us consider how particles move in the ac+dc driven FK model.
The particle motion in the pinning $(F_{\mathsf{dc}}<F_{\mathsf{c}})$  and
the sliding regime ($F_{\mathsf{dc}}>F_{\mathsf{c}}$) is presented in Figure \ref{Fig6} (a) and (b) respectively.

\begin{figure*}[ht] 
\includegraphics[width=15.0cm]{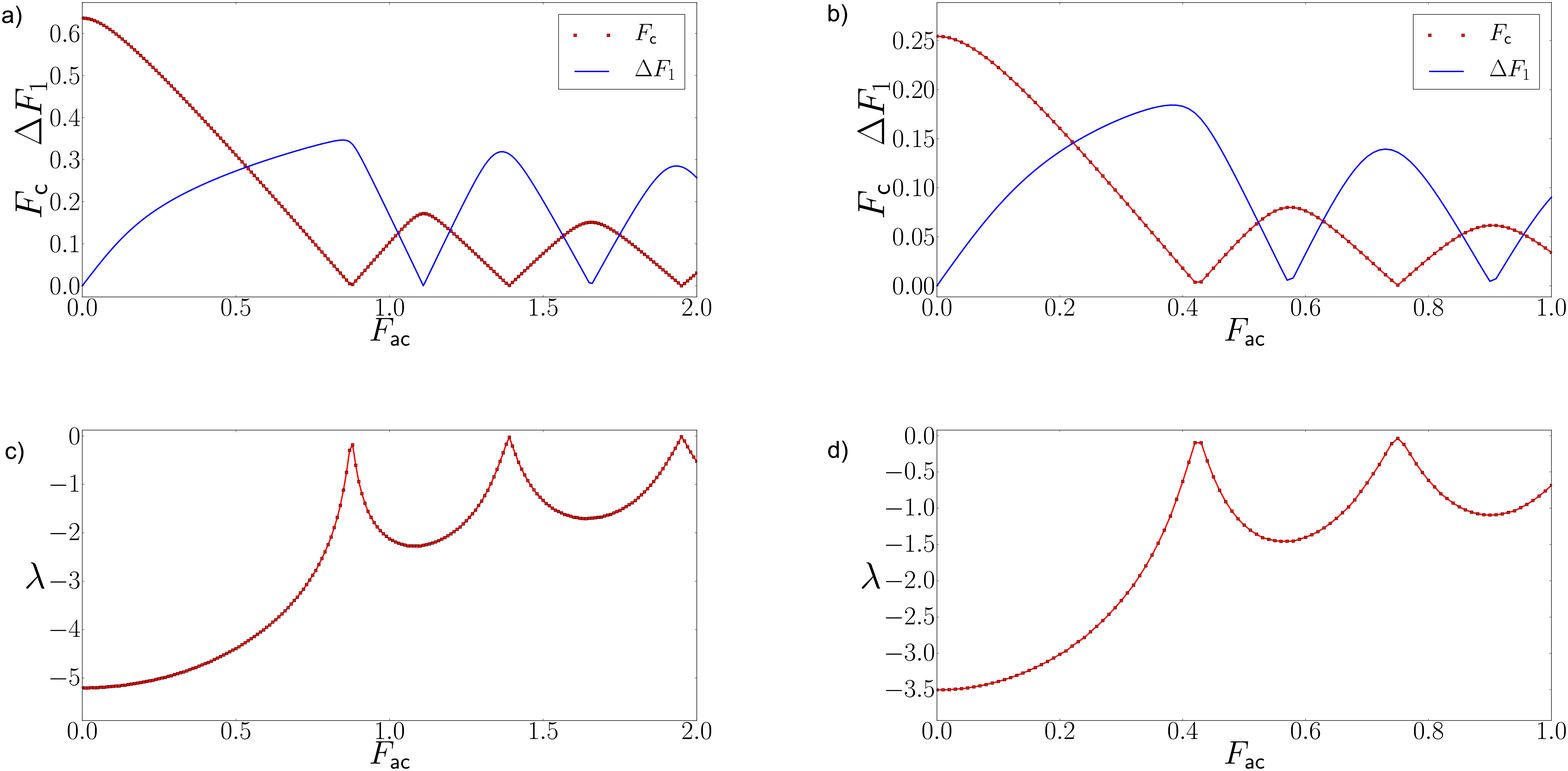}
\centering
\caption{\label{Fig7}
(a),(b) Critical force $F_{\mathsf{c}}$ and harmonic step width $\Delta F_1$ as a function of ac amplitude $F_{\mathsf{ac}}$ for $K=4$,
$\nu_0=0.2$ for $\omega=1$ and $\omega=1/2$ respectively.
Whereas (c),(d) represent Lyapunov exponent $\lambda $ for $F_{\mathsf{dc}}=0$
as a function of ac amplitude for $K=4$, $\nu_0=0.2$ for $\omega=1$ and $\omega=1/2$ respectively.}.
\end{figure*}
\begin{figure*}[ht] 
\includegraphics[width=12.5cm]{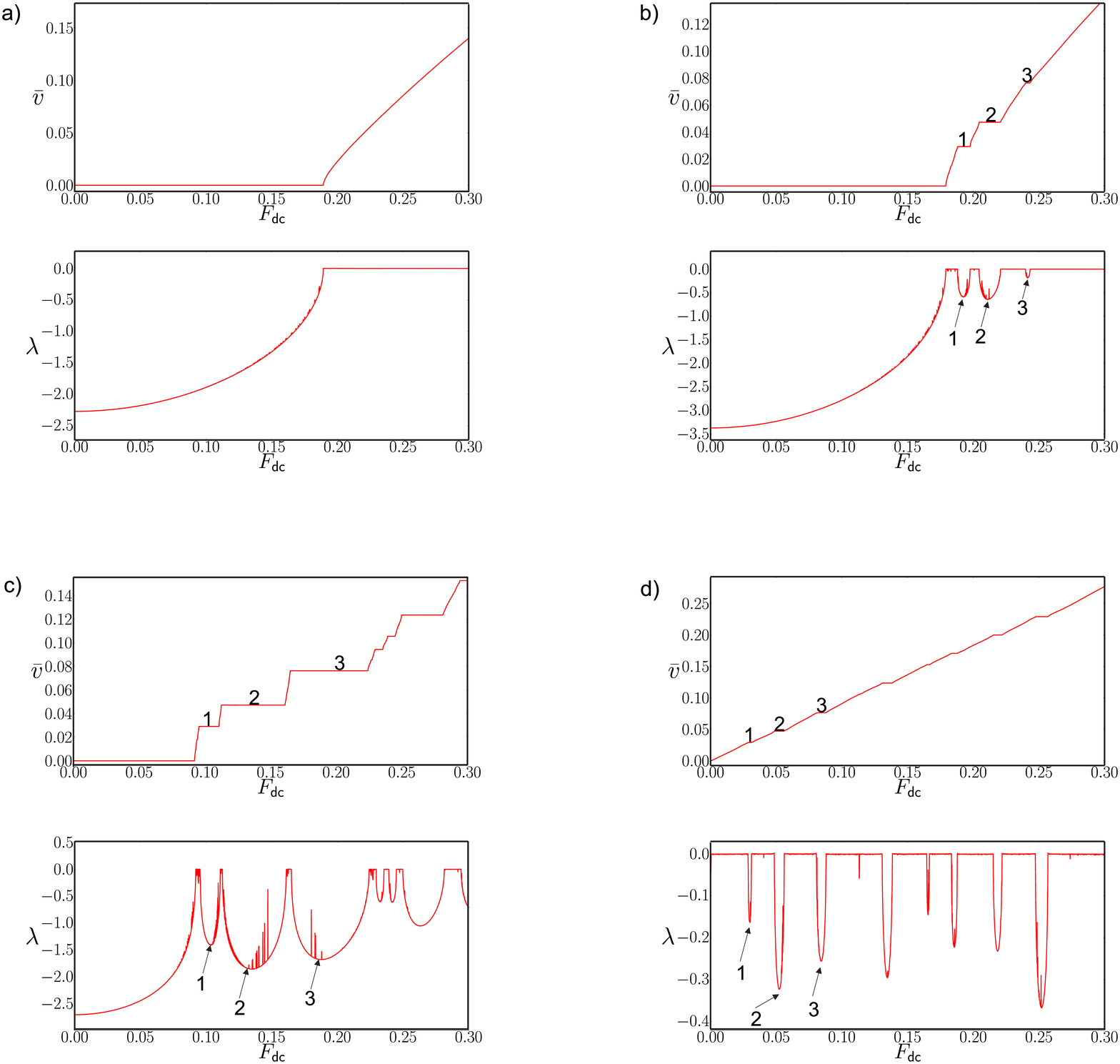}
\centering
\caption{\label{Fig8}
Average velocity and largest Lyapunov exponent as a function of average driving force  for
incommensurate structure $\omega =\frac{89}{144}$, $K=4$, $\nu_0=0.2$ and for different values of
the ac amplitude (a) $F_{\mathsf{ac}}=0$, (b) $F_{\mathsf{ac}}=0.05$, (c) $F_{\mathsf{ac}}=0.2$ and (d) $F_{\mathsf{ac}}=1$.
Number mark first, second and third harmonics.
The unique minimal triplets of integers $(i,j,m)$ in Eq. \ref{vel} that determine the first three harmonic steps are
$(-3,2,1)$, $(2,-1,1)$, $(-1,1,1)$ respectively \cite{Kinezi}.}
\end{figure*}
In the pinning regime the particles and their relative distance to each other  (i.e. first and second particle coordinates $u_1$ and $u_2$ respectively)
are bounded in Figure \ref{Fig6}(a), while on the other hand in the sliding regime in Figure \ref{Fig6} (b) they are not.
The clumping of the data points in the second figure actually points at the ac character of the dynamics
where the particles move for the first part of the period forward and the second one backward but overall they are moving forward.

In general, in the examination of the properties of Shapiro steps, particular attention has been given to their amplitude dependence.

Numerous theoretical and experimental works done in the charge density wave systems \cite{Zettl,Thorn, ThLyoII} and
the systems of Josephson-junction arrays \cite{Fajfer} have been dedicated to the amplitude dependence of the Shapiro steps and the critical depinning force.
It is well known that Shapiro steps width and the critical depinning force exhibit Bessel like oscillations with the amplitude of the ac force
where maxima of one curve corresponds to the minima of the other \cite{Tekic}.
Since LE strongly reflects the presence of the Shapiro steps we will further examine how the ac amplitude affects the LE.

In the usual approach, amplitude dependence of critical depinning force and the step width is determined from the response function
$\bar{\upsilon}(\bar F)$ of the dc+ac driven system where both forces are applied.
On the other hand, in studies of the amplitude dependence of LE, we will consider the pinned state where the dc force is zero $F_{\mathsf{dc}}=0$, and
in the computation of the Lyapunov exponent only the range of $F_{\mathsf{ac}}$ is explored.
The critical depinning force  $F_{\mathsf{c}}$, the size of the first harmonic step $\Delta F_1$ and
the corresponding Lyapunov exponent as a function of the ac amplitude are presented in  Figure \ref{Fig7} for two different commensurate structures.

In Figure \ref{Fig7} (a) and (b), we have typical Bessel-like oscillations of the critical depinning force $F_{\mathsf{c}}$ and
the step size $\Delta F_1$ with the amplitude of ac force.
When compared with these results, the amplitude dependence of LE  in Figure \ref{Fig7} (c) and (d) shows an interesting property.
As we can see, the well known Bessel-like oscillations in Figure \ref{Fig7} (a) and (b) are also reveled
in the amplitude dependence of Lyapunov exponent in (c) and (d).
{\it Lyapunov exponent represents the mirror image of the amplitude dependence for critical depinning force.}
Considering that the LE is determined in the pinning regime without any need to apply dc force and explore dynamical transitions through
a big range of $F_{\mathsf{dc}}$ values for various ac amplitudes in order to get response function and extract the general behavior,
the above results show that LE computational technique might be a very convenient tool for exploring dynamics of driven FK model.
It strongly hints at the Bessel-like nature of the amplitude dependence of critical depinning force and step width without extensive computational effort.

In the case of incommensurate structures, the winding number is irrational.
We will therefore consider the inverse golden mean $\omega=\frac{\sqrt{5}-1}{2}$ \cite{Flor},
which is best approximated by the ratios of two successive Fibonacci numbers, but optimally excluding the first few members of the sequence.
We therefore, use the winding number $\omega=\frac{89}{144}$ in our analysis.
In Figure \ref{Fig8}, the response functions for the  set of parameters $K=4$, $\nu_0=0.2$, $\omega=\frac{89}{144}$
are presented for different values of amplitude of the ac force $F_{\mathsf{ac}}$, where
the insets represents the LE for the same interval of $F_{\mathsf{dc}}$
(the values of parameters are chosen so that system is above Aubry transition \cite{Flor,Kinezi} since
for incommensurate structure below Aubry transition no mode locking is possible \cite{FlorAP}).

As the ac amplitude starts to increase from zero, the steps start to develop, changing in their size and number as
the ac amplitude changes.
The unique minimal triplets of integers $(i,j,m)$ in Eq. \ref{vel} that determine the first three harmonic steps in Figure \ref{Fig8} are
$(-3,2,1)$, $(2,-1,1)$, $(-1,1,1)$ \cite{Kinezi}.
If we analyze only the response function we might think that the large harmonic steps are usually the only one that appear
(some very small subharmonic steps might appear at the high amplitudes).
However, if instead, we calculate the corresponding Lyapunov exponent for the same set of parameters
and intervals of force we can detect the presence of resonances which have been invisible on the response function.
These results in Figure \ref{Fig8} clearly show the advantages of LE computational technique.

Though the LE analysis is applied here to investigate the properties of Shapiro steps it also gives us a possibility to investigate weather the model exibits chaotic behavior.
Presence of chaos has been studied in the Josephson junction systems where even on-step positive Lyapunov exponent was found \cite{Kautz, Shukrinov}.
It is interesting that in our studies of the {\it overdamped} dc+ac driven FK model, even in the case of incommensurate structures and
the large values of ac amplitude $F_{\mathsf{ac}}$, chaos has never been detected.
The main reasons for the absence of chaos are the overdamped character of the model and
the Middleton's no-passing rule \cite{FlorAP, Middleton,Middleton2,Middleton3}.
The rule states that dynamics preserve partial order relation among different configurations, i.e. if $\{u_l(t_0)\}<\{\tilde{u}_l(t_0)\}$
then $\{u_l(t)\}<\{\tilde{u}_l(t)\}$ and it is satisfied in one dimensional dc+ac driven FK model in dissipative limit.
This rule implies that structures do not evolve toward less regularity than they already possess, and that the dissipative character of
the dynamics smooths out any spatial complexity caused by the time evolution.

\section{Conclusion}
\label{concl}

In this paper, dynamics of the driven overdamped FK model is analyzed by using Lyapunov exponent computational technique.
The obtained results have shown that it was often sufficient to calculate the largest Lyapunov exponent
in order to get an insight into the system behavior.
In the dc driven systems, just by looking at the Lyapunov exponent when no driving force is applied we can estimate
how the system will respond to the external force, the more negative Lyapunov exponent is, the more robust system will be.
In the dc+ac driven systems, calculation of LE represents a very convenient tool to detect presence of any dynamical mode-locking or Shapiro steps.
Dependence of the LE on the ac amplitude obtained in the pinned regime represents a mirror image of the amplitude dependance of the critical depinning force.
This clearly shows the advantages of the LE computational technique,
since we can work from the pinned state and without actually driving the system
get an insight into dynamics and the properties of Shapiro steps which appear when dc+ac forces are applied.

The presented results could be important not only for driven dissipative systems such as charge- or spin-density wave conductors and
systems of Josephson junctions which are closely related to the dissipative FK model, but for many areas of physics \cite{FlorAP,Monograph}.
Possibility to estimate the response of the system to the applied force and to get a glimpse into
dynamics just by analyzing one characteristics of the system such as Lyapunov exponent
when the system is static certainly represents an advantage which could help
both theoretical and experimental studies of dynamical systems in general.


\subsection*{Acknowledgment}
We thank Velibor \v Zeli for his help with Matplotlib \cite{Matplotlib}.
This work was supported by the Serbian Ministry of
Education and Science under Contracts No. OI-171009 and No. III-45010 and also by the Provincial Secretariat for High Education and Scientific Research of Vojvodina (project No. APV 114-451-2201).



\end{document}